\documentclass[12pt,preprint]{aastex}
\shorttitle{Coherent Curvature Radiation}
\shortauthors{Asano and Takahara}
\begin{document}
\title{Coherent Curvature Radiation and Proton Counterflow
in the Pulsar Magnetosphere}
\author{Katsuaki Asano}
\affil{Division of Theoretical Astronomy, National Astronomical Observatory of Japan,
              2-21-1 Osawa Mitaka Tokyo, Japan}
\email{asano@th.nao.ac.jp}

\and

\author{Fumio Takahara}
\affil{Department of Earth and Space Science,
            Osaka University, Toyonaka 560-0043, Japan}

\begin{abstract}
In the proton counterflow model of a pulsar magnetosphere that
we have recently proposed,
non-relativistic protons are supplied
from the magnetosphere to flow toward the pulsar surface and 
screen an electric field above the polar cap region.
In this Letter, we show that the proton counterflow 
is also suitable for
the bunching of pair plasma.
The two-stream instability is easily excited 
and can produce bunches of pairs with a relevant length scale
to emit coherent curvature radiation.
\end{abstract}
\keywords{instabilities---plasmas---radiation mechanisms: non-thermal
---stars: pulsars: general}


\section{INTRODUCTION}

\indent

A spinning magnetized neutron star provides huge electric
potential differences between different parts of its surface 
as a result of unipolar induction \citep{gol69}.
A part of the potential difference may be expended on an 
electric field along the magnetic field somewhere in the 
magnetosphere. Although a fully self-consistent model for the 
pulsar magnetosphere has not yet been constructed, several promising 
models have been proposed. Among them, the polar cap model 
\citep{stu71, rud75} assumes that an electric field 
$E_\parallel$ parallel to the magnetic field lines exists just
above the magnetic poles.
The electric field accelerates charged particles up to TeV energies,
and resultant curvature radiation from these particles produces copious 
electron-positron pairs through magnetic pair production.
These pairs are believed to be responsible for radio emission.

The extremely high brightness temperature of pulsar radio emission
requires a coherent source for this radiation.
In the past decade, various maser processes have been proposed and studied:
for example,
curvature emission induced by curvature drift or torsion \citep{luo92,luo95}
and emission due to the cyclotron-Cerenkov and Cerenkov drift instabilities
\citep{lyu99a,lyu99b}.
In most cases these models are sensitive to the magnetic field strength
or other parameters.
Although the torsion-driven maser process \citep{luo95} is not so sensitive
to the physical parameters, we have no definite ideas to achieve
the required inverted energy spectrum of particles and twisted field lines.
Therefore, there is still no widely accepted model
of maser emission.
On the other hand, one of the simplest mechanisms of radiation
is the coherent curvature radiation
from bunches of electron-positron pair plasma,
which was discussed mainly in the 1970s.

If $N$ particles are in a bunch of some characteristic size $\lambda$,
emission with wavelengths longer than $\lambda$ will be coherent.
Then the particles in the bunch
radiate like a single particle with charge $N e$.
The brightness temperature becomes $N$ times that
estimated from the intensity when the $N$ single particles radiate incoherently.

\citet{che77} suggested that the bunching arises from
the electrostatic two-stream instability of electron
and positron streams.
However, \citet{ben77} found that the growth time 
of the instability is far too long.
Although some other mechanisms to excite instabilities
\citep[e.g.][]{gol71,ass80, ass83,uso87}
have been proposed, the problem remains unsolved so far.

Another serious problem in the polar cap model
is the screening of the electric field.
The localized potential drop in the polar cap region
is maintained by a pair of anode and cathode regions.
For a space charge-limited flow model
\citep{faw77, sch78, aro79}, 
where electrons can freely escape from the stellar surface, 
i.e., $E_\parallel = 0$ on the stellar surface,
an anode has been 
considered to be provided by pair polarization.
However, as \citet{shi98, shi02} found,
the thickness of the screening layer is restricted to be
extremely small,
in order to screen the electric field consistently.
A huge number of pairs must be injected within the small thickness.
The required pair multiplication factor per primary particle
is enormously large and cannot be realized in the conventional pair
creation models.

In addition to the above problems, there are various unsolved problems
in pulsar physics; for example, the current closure problem
is a representative one.
It may be worthwhile to explore the possibility of a novel model,
even if the model has some ambiguities at present.
Recently, \citet{asa04} (hereafter AT) have
proposed a new mechanism to screen the electric field.
In the AT model, nonrelativistic protons are supplied
from the corotating magnetosphere to flow toward the stellar surface.
Protons can provide an anode to screen an electric field
the standard polar cap model supposes.
Injected electron-positron pairs yield an 
asymmetry of the electrostatic potential around the screening 
point. The required pair creation rate in this model is 
consistent with the conventional models.

The existence of the  proton counterflows
is also favorable for the bunching of pair plasma.
The existence of protons apparently
makes excitation of two-stream instability easier \citep{che80}.
In this Letter, we show that the pair and proton flows in the AT model
excite electrostatic waves,
which can explain the pulsar radio emission through coherent 
curvature radiation.
In contrast to most modern theories of pulsar radio emission,
the bunch models have been so much discussed from the early days 
of pulsar research and have received several criticisms,
such as insufficient growth rates and rapid bunch dispersion due to 
random motion of bunching particles (see, e.g., Melrose 1995).
However, reconsideration of this old idea without any prejudice
may be valuable \citep{gil04} in order to
overcome present difficulties in pulsar physics,
although the maser models are still an alternative possibility
of the radiation mechanism.
Another interesting reason to reconsider the bunch model is that
space charge density waves appear outside the screening region
in the AT model, even though the velocity dispersion of pairs
is taken into account (see also Shibata et al. 2002).
The two stream instability discussed here may be useful
to excite non-linear radiation processes in a relativistic plasma
\citep{lyu99a,lyu99b}.

In \S 2, we show that the two-stream instability
is easily excited in the AT model.
The excited wavelength is long enough to bunch particles.
In \S 3, we discuss counteraction of the excited waves
on the pair flows.
\S 4 is devoted to a summary and discussion.

\section{TWO-STREAM INSTABILITY}

In this section, we discuss the two-stream instability
for the situation the AT model supposes.
In the AT model, protons are assumed to flow from the corotating magnetosphere
toward the stellar surface.
The primary electron beam is accelerated
from the stellar surface.
Electron-positron pairs start to be injected
at a certain height above the polar cap,
and the electric field is screened there.
Outside this point, two-stream instability may be
excited, and the resultant bunching of pair plasma
yields coherent curvature radiation.

The AT model requires that the proton current $J_{\rm p}$
is of the order of the Goldreich-Julian (GJ)
current $J_{\rm GJ} \equiv -\mathbf{\Omega}_* \cdot \mathbf{B}/
2 \pi$, where $\mathbf{\Omega}_*$ and $\mathbf{B}$ are the angular velocity
of the star and magnetic field, respectively.
Hereafter, we assume $\mathbf{\Omega}_* \cdot \mathbf{B}>0$.
In the case of opposite polarity, the electric fields are not screened
in this model.
There is no observational evidence for two classes of pulsars
due to the polarity, as expected.
The polarity problem remains unsolved so far for all pulsar models.
The proton flow is mildly relativistic
(the average velocity in AT $\simeq -0.4 c$) in the pulsar frame.
The current of the primary electron beam
from the stellar surface is also of the order of $J_{\rm GJ}$.
The Lorentz factor of pairs at injection is required to be more than $\sim 500$.

From the above assumptions,
we obtain the proton number density $n_{\rm p} \simeq \Omega B/
2 \pi c e$.
Then the proton plasma frequency,
$\omega_{\rm pp}^2 \equiv 4 \pi e^2 n_{\rm p}/m_{\rm p}$, becomes
\begin{eqnarray}
\omega_{\rm pp}/2 \pi \simeq 100 T_{0.3}^{-1/2} B_{12}^{1/2} \quad 
\mbox{MHz,}
\end{eqnarray}
where $T_{0.3}$ and $B_{12}$ are the rotation period of the pulsar
and $B$ in units of 0.3 s and $10^{12}$ G, respectively.

In order to simplify the situation, we consider one-dimensional
homogeneous flows of protons and electron-positron pairs.
Since the Lorentz factor of the primary beam of electrons
from the stellar surface
is too large to contribute to the dispersion relation,
we neglect the beam component hereafter.
The distributions of protons and pairs are functions of the 
4-velocity $u=\beta/(1-\beta^2)^{1/2}$, where $\beta=v/c$.
In the linear perturbation theory,
the dispersion relation for electrostatic waves \citep{bal69} 
is given by
\begin{eqnarray}
1+\sum_a \frac{4 \pi q_a^2}{\omega m_a}
\int d u \frac{\beta}{\omega-k v} 
\frac{\partial f_a}{\partial u}=0,
\end{eqnarray}
where $f_a$, $q_a$, and $m_a$ denote the distribution function,
charge, and mass of the particle species $a$, respectively.
Solutions of the dispersion relation usually 
yield a complex frequency $\omega=\omega_r+i \omega_i$.
The imaginary part $\omega_i$ of $\omega$
corresponds to the growth rate of waves.
A positive growth rate $\omega_i>0$ implies an exponentially 
growing wave, while a negative $\omega_i$ implies an 
exponentially damped wave.

In the cold-plasma limit
the distribution function of the proton flow may be written as
$f_{\rm p}=n_{\rm p} \delta(u)$ in the rest frame of the proton flow.
We assume that the number densities and flow velocities of pair electrons and positrons
are the same.
In this case, the contributions of pair electrons and positrons
are degenerate, so that we write the total pair distribution
as $f_\pm=n_\pm \delta(u-u_\pm)$.
We neglect the injection of pairs,
although the pair injection continues outside the polar
cap region in general.
Then we obtain
\begin{eqnarray}
1-\frac{\omega_{\rm pp}^2}{\omega^2}
-\frac{\omega_{\rm p \pm}^2}{(\omega-k v_{\pm})^2 \gamma_\pm^3}=0,
\end{eqnarray}
where $\omega_{\rm p \pm}^2 \equiv 4 \pi e^2 n_\pm/m_e$,
and $\gamma_\pm=500 \gamma_{\pm,5}$ is the Lorentz factor of pairs. 
The charge density wave due to pairs is constructed
by the density waves of electrons and positrons
with displaced phases of $\pi$.

We normalize $k$ and $\omega$ by the proton plasma frequency
as $\tilde{k}=c k/\omega_{\rm pp}$ and
$\tilde{\omega}=\omega/\omega_{\rm pp}$.
We write $\omega_{\rm p \pm} \equiv \xi \omega_{\rm pp}$.
The current of the primary electron beam (we have neglected here)
is also of the order of $J_{\rm GJ}$.
The number of pairs one primary electron creates
may be $10^3$--$10^4$.
Therefore, $n_\pm/n_{\rm p} \equiv M$ may be in the range of $10^3$--$10^4$.
As a result, $\xi=\sqrt{M m_{\rm p}/m_{\rm e}}$ is in the range of 1400--4300.

The condition to yield a pair of complex solutions is given by
\begin{eqnarray}
\tilde{k} \beta_\pm <(1+\zeta)^{3/2},
\end{eqnarray}
where $\zeta \equiv \xi^{2/3}/\gamma_\pm$.
Since $\gamma_\pm$ is required to be larger than $\sim 500$
in the AT model,
we assume $\gamma_\pm=500$--1000.
In this case,
$\zeta$ is in the range of 0.1--0.5, and $\beta_\pm \simeq 1$.
Thus, the threshold of $\tilde{k}$ is close to unity
irrespective of the parameters.

The phase velocity of the excited wave is obtained as
\begin{eqnarray}
\frac{\omega_r}{k} \simeq \frac{v_\pm}{1+\zeta}.
\label{phase}
\end{eqnarray}
The maximum of $\omega_i$ is given
by $\tilde{k}$ smaller than and close to the threshold
$(1+\zeta)^{3/2} \sim 1$.
Since the right-hand side of Eq. (\ref{phase}) is of the order of unity,
$\tilde{\omega_r}$ for the growing wave
is of the order of unity, too.
In general, $\omega_r$ is much larger than
$\omega_i$.
Therefore, $\tilde{\omega}_i$ is about $\sim 0.1$ at most.
Numerical solutions (see Fig. 1) confirm this estimate.

The growth time $\sim 1/(0.1 \omega_{\rm pp}) \sim 10^{-8}$ s $\ll 1/\Omega$,
$R/c$, where $R=10^7 R_7$ cm is the curvature radius of magnetic fields,
is short enough to bunch particles.
Therefore, the bunched particles emit
coherent radiation just above the polar cap.
The wavelength $\lambda$ of the excited wave is  $2 \pi/k
\sim 2 \pi c/\omega_{\rm pp} \sim 300$ cm.
The maximum coherent amplification of curvature radiation
by bunches of pairs is obtained at wavelengths
long compared to the size of bunches $\lambda$.
For wavelengths shorter than $\lambda$ (frequency higher than $\nu_0
\equiv \omega_{\rm pp}/
2 \pi \sim 100$ MHz),
the flux of curvature radiation diminishes with a power law
of some index $\alpha$ as $\propto \nu^{-\alpha}$
\citep[see, e.g.,][]{sag75,ben77,mic82}
extending to the critical frequency $\nu_{\rm c}=3 \gamma_\pm^3 c/4 \pi R
\sim 100 \gamma_{\pm,5}^3 R_7^{-1}$ GHz.
The above frequencies are estimated in the proton rest frame.
Since the proton flow is nonrelativistic,
the redshift of frequencies by the flow is unimportant.
Thus, the coherent radio emission in this model
is well consistent with the observed pulsar radio spectra.

The number of pairs $N$, which can radiate coherently,
may be written as $N \simeq \epsilon M n_{\rm GJ} \lambda^3
\simeq 6 \times 10^{21} \epsilon (M/10^3) (\lambda_3)^3 B_{12} T_{0.3}^{-1}$,
where $\lambda_3=\lambda/300$ cm,
and the factor $\epsilon<1$ is the fraction of particles
that are in a coherent motion.
The cooling time scale
for coherent curvature radiation may be written as
\begin{eqnarray}
t_{\rm cool} \simeq \frac{\gamma_\pm m_{\rm e} c^2}{p N}
\left( \frac{\nu_{\rm c}}{\nu_0} \right)^{4/3}=
\frac{3 R^2 m_{\rm e} c}
{2 e^2 \gamma_\pm^3 N} \left( \frac{\nu_{\rm c}}{\nu_0} \right)^{4/3},
\end{eqnarray}
where $p=2 e^2 c \gamma_\pm^4/3 R^2$ is the emitting power by a single particle.
Here we adopt $\nu_{\rm c}/\nu_0=10^3$.
If $N>3 \times 10^{15}$,
$t_{\rm cool}$ becomes shorter than $R/c \sim 3 \times 10^{-4} R_7$ s.
This is realized for $\epsilon > 5 \times 10^{-7}$ and  
gives a bright enough radio luminosity. Since high  
radio luminosity can be realized even for a smaller $N$ if the number 
of bunches is large enough, this estimate is not unique. 
On the other hand, the radio brightness temperature $T_{\rm b}$ 
may be limited by the self-absorption \citep{che80};
$T_{\rm b}<\gamma_\pm N m_{\rm e} c^2/k_{\rm B}$,
where $k_{\rm B}$ is the Boltzmann constant.
The value $N=3 \times 10^{15}$ gives a limit $\sim 10^{28}$ K,
which is high enough to be consistent with observations.
The above estimate of $N$ and $\epsilon$ appear to be appropriate, 
although it is not unique. Thus, a small value of $\epsilon$ suffices 
for coherent radio emission, although it is hard to estimate $\epsilon$ 
from the first principle
because of some nonlinear effects on the bunching process.

\section{WAVE COUNTERACTION ON THE FLOWS}

The excited waves may produce effective frictional force
on the flows.
If the frictional force is too strong, the force
destroys the structure of the two streams.
We have to construct a model that satisfies the two requirements of
wave excitation in a short timescale and
sustainable structure of the flows, which seems incompatible at first glance.
This problem has not been considered seriously so far.

In this section, we obtain a physical requirement
to make the frictional force negligible.
In the quasi-linear theory, the time-averaged distribution
functions evolve according to \citep[see, e.g.,][]{hin78}
\begin{eqnarray}
\frac{\partial f_a}{\partial t}=\frac{\partial}{\partial u}
\Biggl[ D \frac{\partial f_a}{\partial u} \Biggr],
\end{eqnarray}
where
\begin{eqnarray}
D=\frac{q_a^2}{m_a^2 c^2} {\rm Re} 
\Biggl[\int dk \frac{E_k^2}
{i (k v-\omega_{r,k})+\omega_{i,k}}
\Biggr],
\end{eqnarray}
and $E_k$ is amplitude of the electric field
with a wavenumber $k$.
Then the change of $u$ due to the friction per unit time
is obtained as
\begin{eqnarray}
\dot{u}=\frac{\partial D}{\partial u}.
\end{eqnarray}

The amplitude $E_k$ is difficult to estimate.
The electric field traps pairs,
if the amplitude of the excited electric field $E$ becomes
larger than a threshold
\begin{eqnarray}
E_{\rm max} &\equiv& \frac{k \gamma_\pm m_{\rm e} c^2}{2 e \gamma_{\rm w}^2}
\simeq \frac{\omega_{\rm pp} \gamma_\pm m_{\rm e} c}{2 e \gamma_{\rm w}^2} \nonumber \\
&\simeq& 10^3
\left( \frac{\gamma_{\rm w}^2}{10} \right)^{-1} \gamma_{\pm,5}
B_{12}^{1/2} T_{0.3}^{-1/2} \quad 
\mbox{in esu,}
\label{denba}
\end{eqnarray}
where $\gamma_{\rm w}$ is the Lorentz factor of
the phase velocity of the wave.
We have assumed $\gamma_{\rm w}^2 \gg 1$ in the above estimate.
However, even if $\gamma_{\rm w}^2 \sim 1$,
the correct value differs from the above
at most by a factor of 2.
In the relativistic limit,
$\gamma_{\rm w}^2 \simeq (1+\zeta)/2 \zeta$.
Thus, $\gamma_{\rm w}^2$ is 10 at most.
Assuming the maximum charge density $\rho_{\rm max} \sim E_{\rm max}/\lambda$,
we obtain $\rho_{\rm max} \sim \gamma_\pm m_{\rm e}/(\gamma_w^2 m_{\rm p})
\rho_{\rm GJ}$, where $\rho_{\rm GJ} \equiv \Omega B/(2 \pi c)$ is
the GJ charge density.
Since $\gamma_\pm \sim 10^3$ and $\gamma_{\rm w}^2=1$-$10$,
the maximum charge density turns out to be of the order of
$0.1$-$1$ times the GJ charge density.
Although there may be various nonlinear effects to excite waves,
the maximum amplitude of the electric field may not be much larger
than $E_{\rm max}$.

We approximate the excited electric field
by a monochromatic wave of $E_k=E \delta(k-\omega_{\rm pp}/c)$.
It is apparent that the time scale for the slowdown due to the friction
$t_{\rm fric}$ for the pair flow is longer than that for
the proton flow owing to the large value of $\gamma_\pm$.
Taking into account $\omega_r \simeq \omega_{\rm pp}$ and
$\omega_i \simeq 0.1 \omega_{\rm pp}$,
$t_{\rm fric}$ for the proton flow is obtained as
\begin{eqnarray}
t_{\rm fric}=\frac{1}{|\dot{u}|} \simeq
5 \omega_{\rm pp} \left( \frac{m_{\rm p} c} 
{e E} \right)^2.
\end{eqnarray}
We consider that the physical condition remains unchanged
within a scale $R/c$.
In order to make $t_{\rm fric}$ be longer than $R/c$,
$E$ should be smaller than $E_{\rm max}$ by a factor of 3.
Experience has shown that longitudinal waves are frequently
strongly damped by nonlinear processes.
So we expect that $E$ may be smaller than $E_{\rm max}$,
small enough to keep the fictional force small and 
strong enough to bunch a small fraction of pair particles.

\section{SUMMARY AND DISCUSSION}

We examined the two-stream instability and coherent curvature 
radiation in the proton counterflow model that we have recently proposed. 
The existence of proton flows is favorable
not only for screening of the electric field
but also for the bunching of pair plasma by the two-stream instability.
This model predicts a high growth rate 
and wavelength of electrostatic waves appropriate 
to reproduce observed radio emission by coherent curvature radiation.
The growth rate is basically determined by $\omega_{\rm pp}
\propto \sqrt{B_{12}/T_{0.3}}$.
Since rapidly rotating pulsars tend to have weak magnetic fields,
$\omega_{\rm pp}$ is not so sensitive to the model parameters.
For example, $\omega_{\rm pp}/2 \pi$ becomes about 50 MHz
for $B=10^{9}$ G and $T=1$ ms.
On the other hand, for $B=10^{15}$ G and $T=10$ s, $\omega_{\rm pp}/2 \pi
\sim 500$ MHz.
The resultant growth rate is also insensitive to the model parameters.
It is interesting that the predicted wave length is
comparable to the wave length of the space charge density wave,
which appears outside the screening region in AT.

We noticed the interesting paper of \citet{les98} in which they claimed that
coherent curvature radiation cannot be the source
for the radio emission of pulsars. 
However, their treatment is based on several simplifying assumptions: 
a full coherence extending up to $\nu_{\rm c}$, a large coherence 
volume, and others. Basically, these assumptions lead to 
a short cooling time and a lower value for the  upper limit of luminosity. 
In our case, the coherence volume is smaller and at $\nu_{\rm c}$ 
only a partial coherence is supposed. Moreover, only a small 
fraction of pairs are bunched and the cooling time is 
much longer than their estimate. Thus, the limit claimed by \citet{les98} 
is irrelevant in our case. Examinations of more general constraints 
are beyond the scope of this Letter.

Although the AT model resolves both the screening and radio emission problems,
there remain many ambiguous points:
the frictional force discussed in section 3,
mechanisms to achieve the proton counterflow, and so on.
In the AT model, the proton counterflow comes from the corotating magnetosphere
via the anomalous diffusion \citep{lie85}.
The currents of the primary beam and protons
are most likely to be determined by the global dynamics
in the magnetosphere.
However, we should not persist in the anomalous diffusion,
because there may be other mechanisms to cause the proton counterflow.
Considering that we still do not understand well
the fundamental issues of pulsar physics,
we should be free from any kind of prejudices.
We cannot exclude any possibility at the present stage.

\acknowledgments

This work is  supported in part by a Grant-in-Aid for Scientific Research 
from the Ministry of Education and Science (14079205 and 16540215; F. T.).

\clearpage

\begin{figure}
\epsscale{.70}
\plotone{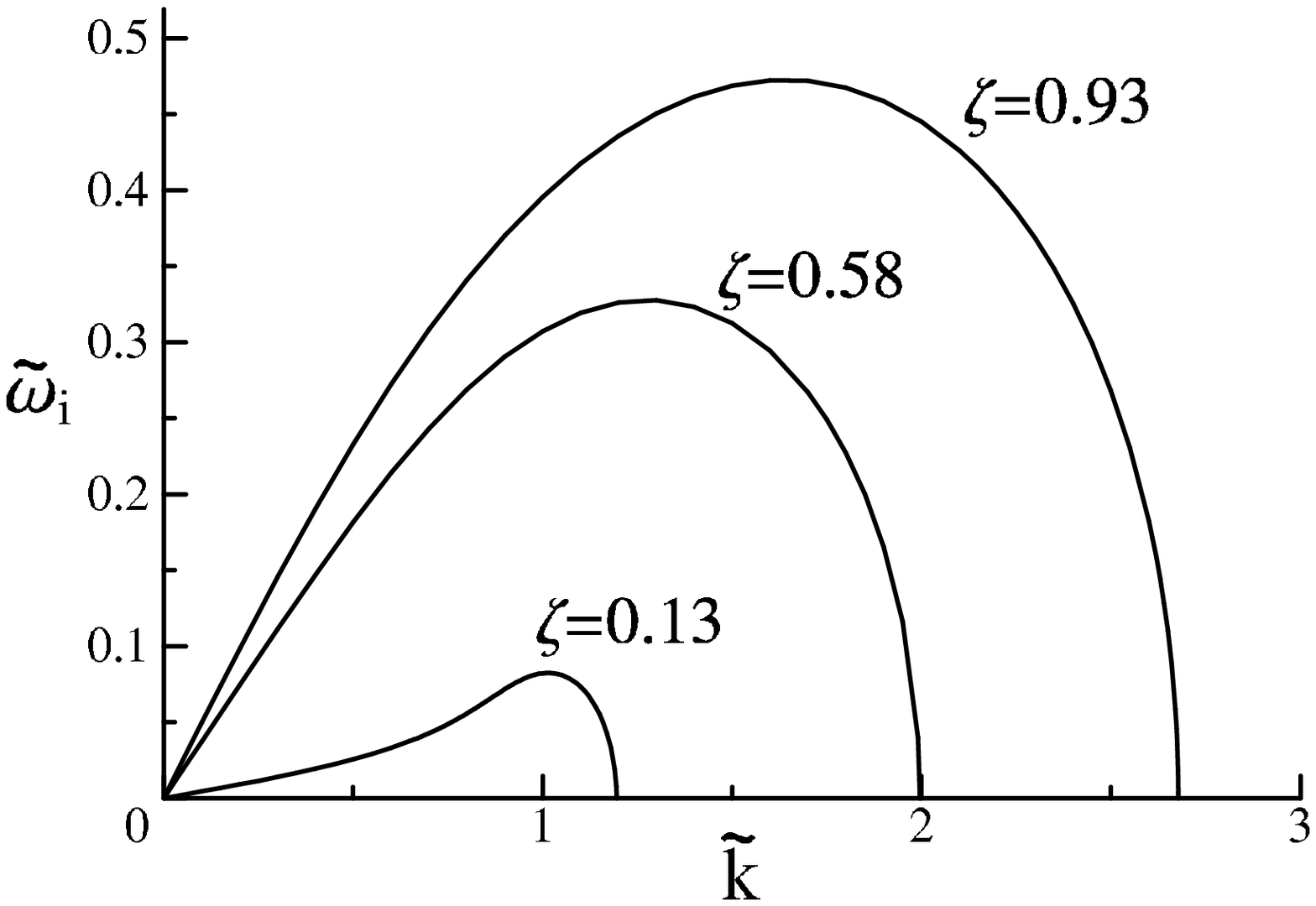}
\caption{Growth rate $\tilde{\omega}_i$ against $\tilde{k}$
for $\zeta=0.13$ ($\xi=500$ and $\gamma_\pm=500$),
$0.58$ ($\xi=5000$), and $0.93$ ($\xi=10^4$).}
\end{figure}

\end{document}